# HRINET: ALTERNATIVE SUPERVISION NETWORK FOR HIGH-RESOLUTION CT IMAGE INTERPOLATION

*Jiawei Li[1], Jae Chul Koh[2], Won-Sook Lee[3]*

[1,3]School of Electrical Engineering and Computer Science, University of Ottawa
[2]Department of Anesthesiology and Pain Medicine, the Korea University Anam Hospital

## ABSTRACT

Image interpolation in the medical area is of high importance as most 3D biomedical volume images are sampled where the distance between consecutive slices is significantly greater than the in-plane pixel size due to radiation dose or scanning time. Image interpolation creates a number of new slices between known slices in order to obtain an isotropic volume image. The results can be used for the higher quality of 3D reconstruction and visualization of human body structures.

Semantic interpolation on the manifold has been proved to be very useful for smoothing image interpolation. Nevertheless, all previous methods focused on low-resolution image interpolation, and most of them work poorly on high-resolution images. We propose a novel network, High Resolution Interpolation Network (HRINet)*,* aiming at producing high quality and realistic CT image interpolations. We combine the idea of ACAI and GANs, and propose a novel idea of alternative supervision method by applying supervised and unsupervised training alternatively to raise the accuracy of human organ structures in CT while keeping high quality. We compare an MSE based and a perceptual based loss optimizing methods for high quality interpolation, and show the tradeoff between the structural correctness and sharpness. Our experiments show the great improvement on $256^2$ and $512^2$ images quantitatively and qualitatively.

## 1. INTRODUCTION

Since most 3D biomedical volume images are sampled at locations where the distance between consecutive slices is significantly larger than in-plane pixel size [13][14][15]. Especially, the number of scans during in-vivo CT data scanning is limited in order to avoid the risks of radiation on patients [16], and large gaps between slices might lead to bias during the process of 3D reconstruction of bone geometry and tissue features [17]. Image interpolation is used to create a number of new slices between known slices in order to obtain an isotropic volume image. The results can also be used to visualize the CT data in any view including sagittal, coronal and axial views regardless in which view the scanning is done [17][18].

To get accurate and semantic smooth interpolations between slices, we must know the manifold of latent representation of them. The autoencoder [19] has been proved to be efficient for dimensionality reduction generating latent representation of images [20][21][22]. The highly abstract and entangled latent code captures the semantic representation of images on the manifold, and linearly interpolation between latent space reflects the semantic smooth interpolation on data space [20][23]. The ACAI model [23] show the potential of having autoencoder learn the latent representing way correctly under the adversarial constraints. Even though the interpolated results are clear, interpolating between high-resolution images is always a very challenging task. Because the autoencoder based models are all have a common defection: The autoencoder is lossy way for image compression, although it can correctly capture the latent representation, rough shape and structure can be reconstructed by decoder, fine details like texture are reluctant to be recovered, especially observable in the case of highly dimensionality reduction [20][24]. And the size and dimensionality of the latent code are very sensitive for interpolation. Inappropriate size of latent code might lead to overlapping or blurry problem [23] for interpolated image as illustrated in **Figure 1**. Concretely, the overlapping problem would happen when the dimensions of latent code is too much relative to image size. And the blurry case is usually caused by the too few dimensions for feature representation.

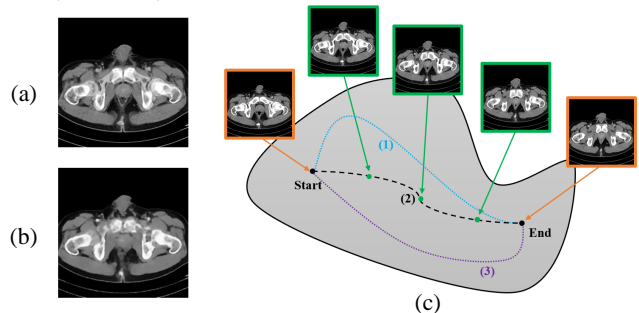

**Figure 1**: Example of overlapping and blurry problems. (a) The interpolated result is sharp but overlapped, close to interpolating in data space rather than latent space. (b) Interpolating in latent space, but the result is blurry and lacks fine details. (c) "Appropriate" interpolations on the manifold. There are many possible paths like (1) (2) and (3) connect a pair of endpoints, but only (2) is considered semantically correct.

## 2. RELATED WORK

### 2.1. Generative Adversarial Networks

Generative Adversarial Networks (GANs) [1][2] have achieved great success among many image generation related tasks, including representation learning [2], image super-resolution [3][4] and neural style transfer [5][6], etc. GANs take unsupervised adversarial training, by minimizing Jensen-Shannon divergence between data distribution $\mathbb{P}_r$ and model distribution $\mathbb{P}_g$. The overall the minimax objective for optimization is:

$$\min_G \max_D \mathbb{E}_{x \sim \mathbb{P}_r}\left[\log(D(x))\right] + \mathbb{E}_{x \sim \mathbb{P}_g}\left[\log(1 - D(\tilde{x}))\right] \quad (1)$$

The GANs are hard to train as the adversarial balance can easily be broken during training, which often leads to vanishing gradient problems in Generator [7]. Besides, the "model collapse" is another critical issue for GANs. DCGAN [2] explored a stable network architecture for training. Wasserstein GANs (WGAN) [7] proposed the wasserstein distance loss, which is the minimum cost of transporting mass between $\mathbb{P}_r$ and $\mathbb{P}_g$ instead of using the JS divergence. Gulrajani et al. [8] introduced gradient penalty to WGAN and further improved the quality of generated images. Other people like Mao et al. [9], Miyato et al. [10] also made great contributions to the improvement of GANs.

## 2.2. Autoencoders and interpolation

Autoencoders [19] contains encoder and decoder, both of which nowadays refer to multi-layer neural networks. Autoencoder has shown great potential of creating the latent representation of input. Interpolating using an autoencoder refers to the process of using the decoder $De_\varphi$ to decode a mixture of two latent codes produce by the corresponding encoder $En_\theta$. Previous works [20][33][34] showed that all of vanilla autoencoder, Variational Autoencoder (VAE), Vector Quantized Variational Autoencoder (VQ-VAE) and denoising autoencoder have ability for interpolation. But they have blur, distortion or semantics unsmooth problems.

Adversarially Constrained Autoencoder Interpolation (ACAI) addresses above issues by adding adversarial regularizer and extra discriminator. The discriminator tries to predict the interpolation coefficient α from interpolated results, and the autoencoder is trained to fool the discriminator to think that α is always zero. The optimization objective could be expressed as:

$$\mathcal{L}_d = ||d_\omega(\hat{x}_\alpha) - \alpha||^2 + ||d_\omega(\gamma x + (1-\gamma)g_\phi(f_\theta(x)))||^2 \quad (2)$$
$$\mathcal{L}_{f,g} = ||x - g_\phi(f_\theta(x))||^2 + \lambda ||d_\omega(\hat{x}_\alpha)||^2 \quad (3)$$

where $\gamma$ is a scalar hyperparameter in the range of [0, 1], and $\lambda$ controls the weight of the regularization term.

## 3. HIGH-RESOLUTION INTERPOLATION

The diagram of our proposed model is shown in **Figure 2** (a). The encoder is trained to acquire the latent space representation of input CT slices, which are randomly chosen from the dataset.

### 3.1. Interpolation between feature maps

The interpolation coefficient α is sampled from the uniform distribution $\alpha \in \mathcal{U}(0, I)$. Differently from ACAI model [23] that did only the linear interpolation among latent code, we add the multidimensional interpolation between feature maps in lower layers. We make lower layer feature maps capture more generic features of the inputs, like rough shape or topology, and higher layers of feature maps contain fine details, like tissues or organ edges. Compared with ACAI, our model delivers more information hierarchically from encoder to decoder and suppress the occurrence of the overlapping and blurry problem described previously.

The decoder combines mixed feature maps and latent code in lower layers and disentangle the latent representation to generate visually pleasing results. Thus, the bottom-up reconstruction strategy of features maps can recover more details from the highly abstract representation in the premise that the interpolated latent code is acquired on the manifold. **Figure 2** (b) is the architecture of the autoencoder part in the model.

### 3.2. PatchGAN regularizer

In our model, two discriminators *(D1, D2)* are used for regularizing the Generator (*G*) to generate high-resolution and smoothly interpolated results. *D1* proposed by ACAI [23] is used to predict the interpolation coefficient α from the mixing of latent code. And the *G* is trained to fool the *D1*, making the *D1* output α = 0 consistently regardless of the input.

Second discriminator *D2* that is a Markovian discriminator, PatchGAN, proposed by Zhu et al. [6][27] determines whether every patch of image is real or not. Taking PatchGAN discriminator to be the regularizer can also promote the generator to generate images with more fine details and sharp edges. The superiority of PatchGAN over the counterpart of traditional discriminator is that it focuses on local response of images rather than the global one, which is good for modeling only high-frequency components [6], whereas low-frequency counterpart is restricted by L2 loss term among the autoencoder part. Thus, suffering less impact from whole image semantics makes PatchGAN to be a better details and texture regularized discriminator.

### 3.3. Alternatively supervising training

It is unavoidable that the generator tends to generate images with incorrect shape or details as illustrated in **Figure 1** (c) and it's very hard to control in the latent space because the generator does not know which path to take is considered "correct" by human level. To address that issue, we regularize the generator to train with the real dataspace, to correct the "deviation to the inappropriate path" and force it to the semantic meaningful path on the manifold. We make the generator train in two different stages: the unsupervised training step (Step I) and supervised training stage (Step II).

*3.3.1 Unsupervised training step (Step I)*

In Step I, the generator is trained with $D_1$ and $D_2$. We use $En_\theta$, $De_\phi$ to denote the encoder (parametrized by $\theta$) and decoder (parametrized by $\phi$) within *G*, the encoder and decoder are trained simultaneously with some kinds of loss function. For a given interpolation coefficient α ∈ [0,1], the convex combination of latent code can be written by: $\hat{z}_\alpha = \alpha En_\theta(x_1) + (1-\alpha)En_\theta(x_2)$. Therefore, the interpolated result is $\hat{x}_\alpha = De_\phi(\hat{z}_\alpha)$.

The overall objective for Step I is:

$$\mathcal{L}_{ae} = \underbrace{\mathcal{L}_{cont}}_{\text{content loss}} + \lambda \underbrace{\mathcal{L}_{ac}}_{\text{adversarial constrain loss}} + \beta \underbrace{\mathcal{L}_{Gen}}_{\text{adversarial loss}} \quad (4)$$

For content loss, it can be calculated by pixel-wise MSE loss [7][29] or perceptual loss [3][30][31], calling the model either as **HRINet**-MSE, the same as the corresponding item in ACAI [23] or as **HRINet**-Perceptual respectively. Instead of using the standard pre-trained VGG net [32] for perceptual loss calculation, which is commonly used in natural image super-resolution [3][30], we use the pre-trained encoder itself (pre-trained in Stage I, see **Section 4.2**) to be a network for loss calculation. The network is fixed during training and only for calculating perceptual loss denoted as $P_\varphi$ (parametrized by $\varphi$).

$$\mathcal{L}_{cont} = \underbrace{\left|\left|x - De_\phi\big(En_\theta(x)\big)\right|\right|^2}_{\text{MSE content loss}} \quad (5)$$

or

$$\mathcal{L}_{cont} = \underbrace{||P_\varphi(x) - P_\varphi(De_\phi\big(En_\theta(x)\big))||^2}_{\text{perceptual content loss}} \quad (6)$$

Note that during this step, the content loss can be regarded as reconstruction error where $x$ and $De_\phi\big(En_\theta(x)\big)$ refer to the input image and the reconstructed result. Step I is still be considered unsupervised training step relatively because the generated interpolation $\hat{x}_\alpha$ are only for the purpose of calculating regularization items $\mathcal{L}_{ac}$ and $\mathcal{L}_{Gen}$ instead of being used to minimize the distance between ground truth as illustrated in **Section 3.3.2**.

For adversarial constrain loss, we keep the same loss item in ACAI, which purpose is trying to fool the *D1* (parametrized by $\omega$) to think that the input is always a non-interpolated one.

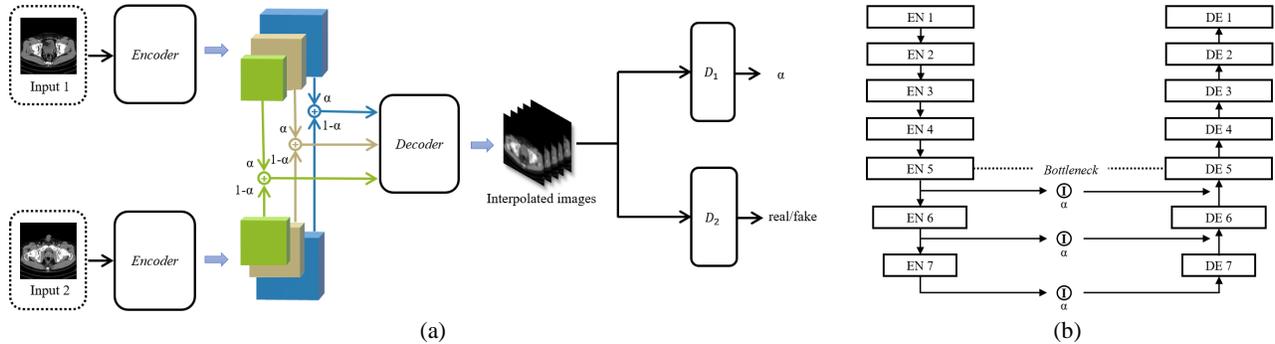

**Figure 2**: (a) The overview of our model. An encoder is trained to produce low-level feature maps which are interpolated with interpolation coefficient $\alpha$, and the decoder tries to disentangle the latent representation and mapping it into data space. $D_1$ is the discriminator tries to predict $\alpha$, and $D_2$ is trained to distinguish whether the input is real or not. The overall autoencoder can be regarded as Generator, trained to making the $D_1$ output $\alpha = 0$ consistently. (b) Network architecture of the autoencoder part. "EN" or "DE" stands for encoder or decoder block. The low-level feature maps interpolation is done beyond the autoencoder part, and α is interpolation coefficient. Bottleneck is where the numbers of channels getting decreased dramatically.

$$\mathcal{L}_{ac} = ||D_{1_\omega}(\hat{x}_\alpha)||^2 \quad (7)$$

The adversarial loss is common generative loss in GANs, making $G$ generate indistinguishable interpolated results for $D_2$ (parametrized by $\psi$), so as to further encourage the network to favor solutions that reside on the manifold.

$$\mathcal{L}_{Gen} = log\left(1 - D_{2_\psi}(\hat{x}_\alpha)\right) \quad (8)$$

*3.3.2 Supervised training step (Step II)*

In Step II, we customize every input batch to be a group of $n$ consecutive CT slices chosen from the same patient. For each patient, the longitudinal distance between slices is a known constant.

The first and last slices within the group would be two inputs images for interpolation, which are a pair of endpoints in dataspace, the semantic distance on the manifold is relatively farther (compared with two adjacent slices for interpolation), we describe the case as "interpolating on a large gap". The generator is made to generate $n - 2$ interpolated images (endpoints are excluded) and $\alpha$ is equally spaced from $(0, 1)$. Then we try to minimize the distance between the interpolated result and corresponding real slices to eliminate the unrealistic features. The optimization objective in Step II is:

$$\mathcal{L}_{ae} = \frac{1}{n-2}(\underbrace{||x - \hat{x}_\alpha||^2}_{\text{MSE content loss}} \text{ or } \underbrace{||P_\varphi(x) - P_\varphi(\hat{x}_\alpha)||^2}_{\text{perceptual content loss}}) \quad (9)$$

depending on whether it is **HRINet**-MSE or **HRINet**-Perceptual.

Typically, the further apart the two endpoints are on the manifold, the harder for the model to create semantic interpolation smoothly between them. The purpose of training the model to interpolate on a large gap is to improve the robustness and stability of the model.

These two steps run alternatively. The regularization effect of both steps all has a great impact on the generator. Besides, the alternative training process shows the potential to combine two or more totally different training steps, that means several optimization methods could be applied simultaneously without conflict. It has the same effect on including all training steps in once back propagation but might become impractical due to occupying too much memory.

## 4. EXPERIMENTS

### 4.1. Dataset Description

The Dataset contains around 1400 CT slices from 10 different patients provided by the *Korea University Anam Hospital*. The patients are among different ages and gender. All slices are taken from the torso cross section, including abdomen, pelvis, etc. The slice thickness is from *3mm* to *5mm*. The original data is resized into $512^2$ for the purpose of training and testing with a ratio of 9:1.

### 4.2. Training details and parameters

The model is parameter sensitive and hard to train at once. To speed up the convergence, we divided our training process into two stages:

In stage I, we only trained the autoencoder with discriminator $D_1$ for 50000 iterations, in order to make the generator have ability to generate interpolated results but in low quality. In stage II, we used the pre-trained network in stage I, and made it train with discriminators $D_1$ and $D_2$. We set generative adversarial coefficient $\beta = 5 \times 10^{-4}$ for $256^2$ interpolation and $\beta = 5 \times 10^{-3}$ for $512^2$ interpolation. We also found that large $\beta$ might mislead the gradient descent direction of $G$ that resulted in the whole training collapse. In Step II "interpolating on a large gap", we set $n = 7$, and 5 intermediate slices are used for content loss calculation. Because the model is robust enough and able to satisfy most interpolation tasks in reality under the training of large gaps that span six times of slice thickness. Any interpolation results whose endpoints distance less than six times of slice thickness during our training would be considered confidential. The Stage II training process runs for the same number of iterations as Stage I.

During the two stages, we kept the learning rate to be $1 \times 10^{-4}$ and adversarial constrain coefficient $\lambda = 0.5$ consistently. The size of feature maps (latent code) is shown in **Table 1**. In our model, we make features be mixed in a more hierarchical way.

**Table 1**: The contrast of feature map sizes be passed through for interpolation between ACAI and HRINet.

| Network | ACAI | | **HRINet**(Ours) | |
|---|---|---|---|---|
| Input size ($w \times h \times c$) | 256×256×1 | 512×512×1 | 256×256×1 | 512×512×1 |
| Feature map size ($w \times h \times c$) | 16×16×16 | 32×32×16 | 16×16×16 | 32×32×16 |
| | - | - | 8×8×8 | 16×16×8 |
| | - | - | 4×4×8 | 8×8×4 |

### 4.3. Evaluation

We evaluate the performance of our models without adversarial components ($D_1$ and $D_2$) and the effect of two kinds of loss applied in Stage II alternatively training process. For contrast, if neither of the above is applied, that is, only the PatchGAN ($D_2$) is added to the original model and the alternative training part is removed.

Quantitative results are summarized in **Table 2** and some of visual examples are provided in **Figure 4** and **Figure 5**. **Figure 4** shows the qualitative comparison and locally enlarged view. Both PSNR and SSIM are calculated for specific interpolated results which are intercepted from a series of continuous interpolations. PSNR and SSIM quantitate measure the distortion and structural similarity between the results and originals. However, both quality metrics can hardly represent high-frequency components and sometimes deviate from human perception. MSE based solutions lead to high scores but often result in overly smoothing effects [3][28]. It corresponds to the results in **Figure 4**, where ACAI and HRINet-MSE has spatial MSE loss items and they get higher PSNR/SSIM scores, but clearly much blurrier than HRINet-Perceptual. We observe that there's a tradeoff between the structural correctness and sharpness.

Moreover, **Figure 4** also shows the benefits of having the model get access to the real data distribution, no matter in MSE loss way or perceptual loss way, and the concrete implementation is discussed in **Section 3.3**. We observe that in large gap interpolation tasks (when the interpolation gap is six consecutive slices), both ACAI and HRINet (PatchGAN only) cannot correctly "guess" the "trend of bone shape changing" which result in poorly interpolations. In other words, the model unable to find a semantic meaningful path to go through between two endpoints of the manifold.

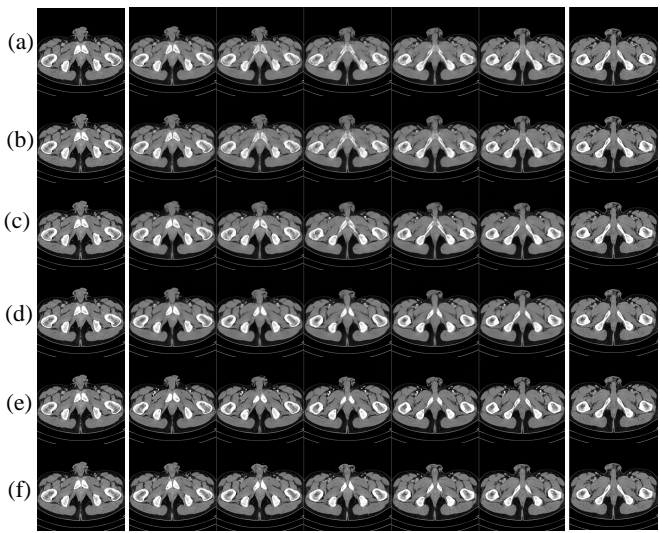

**Figure 3**: Example interpolations on a large gap where ground truth is available. The leftmost and rightmost columns are known slices as inputs, and the gap between them is five consecutive slices. Results are produced by (a) AE, (b) ACAI, (c) HRINet (PatchGAN only), (d) HRINet-MSE, (e) HRINet-Perceptual, (f) Ground truth.

In **Table 2**, we calculate average PSNR/SSIM, RMSE and Ma et al.[11] scores for each of the model respectively over one hundred of images. The results show that HRINet-MSE and HRINet-Perceptual often lead to the best or the second best scores. As expected, the models with spatial MSE loss item (ACAI and HRINet-MSE) perform well in pixel-wise distance metrics (PSNR/SSIM and RMSE), while HRINet-Perceptual and HRINet (PatchGAN only) do well in no-reference perceptual similarity metrics. And perceptual based metrics like Ma et al.[11] has been proved to be closer to human visual judgement.

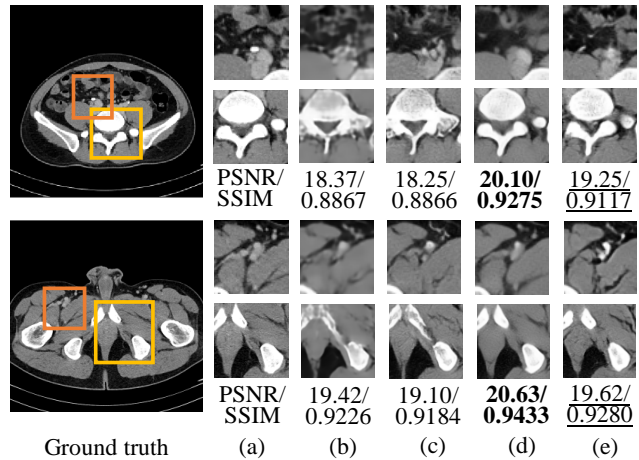

| | | (a) | (b) | (c) | (d) | (e) |
|---|---|---|---|---|---|---|
| | PSNR/SSIM | | 18.37/0.8867 | 18.25/0.8866 | **20.10/0.9275** | 19.25/0.9117 |
| | PSNR/SSIM | | 19.42/0.9226 | 19.10/0.9184 | **20.63/0.9433** | 19.62/0.9280 |
| Ground truth | | | | | | |

**Figure 4**: Qualitative results and locally enlarged view. The upper left and lower left are the ground truth of interpolations, other results are intercepted from a series of continuous interpolations. The best result is highlighted in bold and the second best is underlined. (a) Ground truth, (b) ACAI, (c) HRINet (PatchGAN only), (d) **HRINet**-MSE, (e) **HRINet**-Perceptual.

## 5. CONCLUSION

In this paper, we propose a novel way of creating smooth CT interpolations with high quality, which is very useful to fill the gaps between adjacent CT slices. This is achieved by HRINet combining two phases of the training process, the unsupervised pre-training and alternatively supervising training. We show our results are superior to other existing interpolation approaches such as autoencoder or ACAI quantitatively and qualitatively. HRINet has made improvements in both accuracy and sharpness for CT slices interpolation, and we also find the tradeoff between HRINet-MSE and HRINet-Perceptual in terms of the criterion of structural correctness and sharpness. However, there is still room for further improvement for our model. One of the drawbacks is that some details like tissues and vessels in the CT images are not properly interpolated compared with skeletons and organs, which needs to be explored further.

**Table 2**: Different metrics for evaluating the performance of our model. The best result is highlighted in bold and the second best is underlined. (PSNR/SSIM and Ma et al. [11] are the higher the better, RMSE is the lower the better.)

| *Metrics* | Size | AE | ACAI | HRINet (PatchGAN only) | **HRINet**-MSE | **HRINet**-Perceptual | Ground truth |
|---|---|---|---|---|---|---|---|
| *PSNR/SSIM* | $256^2$ | 22.730/0.957 | 22.800/0.957 | 22.130/0.951 | **25.576/0.977** | 22.654/0.958 | $\infty/1.0$ |
| | $512^2$ | 21.937/0.907 | 21.968/0.909 | 20.141/0.872 | **21.824/0.909** | 20.826/0.894 | $\infty/1.0$ |
| *RMSE* | $256^2$ | 16.985 | 16.845 | 18.072 | **12.059** | 16.6326 | 0 |
| | $512^2$ | 18.394 | **18.290** | 22.435 | 18.305 | 20.4879 | 0 |
| *Ma et al.[11]* | $256^2$ | 2.811 | 2.843 | 2.882 | 2.856 | **2.883** | 2.950 |
| | $512^2$ | 2.601 | 2.577 | 2.893 | 2.934 | **2.961** | 2.911 |